\documentclass[11pt]{article}
\usepackage{charter} 
\usepackage{fullpage}
\usepackage{titlesec}
\usepackage[hyphens]{url}     
\usepackage[hidelinks]{hyperref}       
\usepackage{booktabs}       
\usepackage{amsmath,graphicx}
\usepackage{amssymb,psfrag,mathtools}
\usepackage{cite}
\usepackage{url} 
\usepackage{float}
\usepackage{color}
\usepackage{bm}
\usepackage{bbm}
\usepackage{algorithm}
\usepackage{algpseudocode}
\usepackage{enumitem}
\usepackage{siunitx}
\usepackage{multirow}
\usepackage{units}

\algtext*{EndIf}
\algtext*{EndFor}
\algtext*{EndWhile}

\def\defn{\,\coloneqq\,}
\def\argmin{\mathop{\mathrm{arg\,min}}} 
\def\zer{\mathrm{zer}} 

\def\ebm{{\bm{e}}}

\def\xbm{{\bm{x}}}

\def\ybm{{\bm{y}}}
\def\zbm{{\bm{z}}}

\def\vbm{{\bm{v}}}

\def\zerobm{\bm{0}}

\def\Abm{{\bm{A}}}

\def\Wbm{{\bm{W}}}

\def\xbmast{{\bm{x}^\ast}}

\def\xbmhat{{\widehat{\bm{x}}}}

\def\Tsf{{\mathsf{T}}}

\def\Dsf{{\mathsf{D}}}

\def\Gsf{{\mathsf{G}}}
\def\Isf{{\mathsf{I}}}

\def\Rsf{{\mathsf{R}}}

\def\R{\mathbb{R}}

\begin{document}

\title{Monotonically Convergent Regularization by Denoising}

\author{
Yuyang~Hu$^{1}$, Jiaming~Liu$^{1}$, Xiaojian~Xu$^{2}$,~and~Ulugbek~S.~Kamilov$^{1, 2}$\\
\emph{\footnotesize $^{1}$Department of Electrical and Systems Engineering,~Washington University in St.~Louis, MO 63130, USA}\\
\emph{\footnotesize $^{2}$Department of Computer Science and Engineering,~Washington University in St.~Louis, MO 63130, USA}
}

\date{}

\maketitle 

\begin{abstract}
Regularization by denoising (RED) is a widely-used framework for solving inverse problems by leveraging image denoisers as image priors. Recent work has reported the state-of-the-art performance of RED in a number of imaging applications using pre-trained deep neural nets as denoisers. Despite the recent progress, the stable convergence of RED algorithms remains an open problem. The existing RED theory only guarantees stability for convex data-fidelity terms and nonexpansive denoisers. This work addresses this issue by developing a new \emph{monotone RED (MRED)} algorithm, whose convergence does not require nonexpansiveness of the deep denoising prior. Simulations on image deblurring and compressive sensing recovery from random matrices show the stability of MRED even when the traditional RED algorithm diverges.
\end{abstract}

\section{Introduction}
\label{sec:intro}

Many imaging problems, such as deblurring and super-resolution, can be formulated as \emph{inverse problems} involving the recovery of an image $\xbm \in \R^n$ from its measurements
\begin{equation}
\label{Eq:ForwardProblem}
\ybm = \Abm\xbm + \ebm,
\end{equation}
where $\Abm \in \R^{m \times n}$ is the measurement operator and $\ebm \in \R^n$ is the noise. Since many inverse problems are ill-posed, it is common to formulate the solution as a \emph{regularized inversion}, expressed as an optimization problem
\begin{equation}
\label{Eq:RegularizedInversion}
\xbmhat = \argmin_{\xbm \in \R^n} f(\xbm) \quad\text{with}\quad f(\xbm) = g(\xbm) + h(\xbm),
\end{equation}
where $g$ a \emph{data-fidelity term} that quantifies consistency with the observed measurements $\ybm$ and $h$ is the \emph{regularizer} that enforces prior knowledge on $\xbm$. For example, when the noise vector $\ebm$ in~\eqref{Eq:ForwardProblem} is the \emph{additive white Gaussian noise (AWGN)}, the popular data-fidelity term is the least-squares function ${g(\xbm) = \frac{1}{2}\|\ybm-\Abm\xbm\|_2^2}$. On the other hand, many widely-used image regularizers are based on sparsity promoting functions of the form $h(\xbm) = \tau\|\Wbm\xbm\|_1$, where $\tau > 0$ is the regularization parameter and $\Wbm$ is a suitable transform~\cite{Rudin.etal1992, Bioucas-Dias.Figueiredo2007, Beck.Teboulle2009a}. 

There has been considerable recent interest in leveraging image denoisers as priors for the recovery of $\xbm$~\cite{Venkatakrishnan.etal2013, Schniter.Rangan2015}. One of the most popular frameworks in this context is \emph{regularization by denoising (RED)}~\cite{Romano.etal2017}, which uses a denoiser $\Dsf: \R^n \rightarrow \R^n$ as a regularizer within an iterative algorithm. For example, the \emph{steepest descent} variant of RED (RED-SD)~\cite{Romano.etal2017} is implemented by iterating the following steps until convergence
\begin{subequations}
\label{Eq:REDSD}
\begin{align}
\label{Eq:REDSD1}&\xbm^k \leftarrow \xbm^{k-1} - \gamma \Gsf(\xbm^{k-1})\quad\text{with}\\
\label{Eq:REDSD2}&\quad \Gsf(\xbm^{k-1}) \defn \nabla g(\xbm^{k-1}) + \tau (\xbm^{k-1} - \Dsf(\xbm^{k-1})),
\end{align}
\end{subequations}
where $\gamma > 0$ is the step size, and $\tau > 0$ is the regularization parameter. For a locally homogeneous $\Dsf$ that is nonexpansive and has a symmetric Jacobian, it can be shown~\cite{Romano.etal2017, Reehorst.Schniter2019} that RED algorithms solve~\eqref{Eq:RegularizedInversion} with a regularizer
\begin{equation}
\label{Eq:REDRegularizer}
{h(\xbm) = \frac{\tau}{2}\xbm^\Tsf(\xbm-\Dsf(\xbm))}.
\end{equation}
The state-of-the-art performance of RED on many computational imaging problems has motivated a great deal of research on RED and its applications~\cite{Metzler.etal2018, Sun.etal2020a, Mataev.etal2019, Cohen.etal2021, Liu.etal2020, Liu.etal2021b}.

Despite the recent activity around the topic, the stable convergence of RED algorithms remains an open problem from both theoretical and practical perspectives. The existing theoretical guarantees for the convergence of RED algorithms require the denoiser $\Dsf$ to be a \emph{nonexpansive} operator~\cite{Romano.etal2017, Sun.etal2020a, Reehorst.Schniter2019, Cohen.etal2021, Liu.etal2021b}. On the other hand, it has been empirically observed that RED algorithms can easily diverge for expansive denoisers (see, for example, Table~II in~\cite{Sun.etal2020a}). With the explosion of interest in using deep neural networks within iterative algorithms, there is consequently a pressing need for more flexible strategies for desining RED algorithms that are stable for nonconvex functions $g$ and expansive operators $\Dsf$.

This paper addresses the need for stable RED algorithms by proposing a new \emph{monotone RED (MRED)} algorithm that can offer stable convergence for nonconvex data-fidelity terms and expansive deep image denoisers. MRED defines and uses an explicit loss based on the fixed-point interpretation of RED that does not rely on the existence of the explicit RED regularizer $h$. MRED extends the heuristic \emph{backtracking line search (BLS)} strategy proposed in the prior work~\cite{Liu.etal2020} by adopting a more principled BLS strategy. Our simulations on image deblurring and compressive sensing from random matrices illustrate the stability of MRED compared to the traditional RED-SD with and without BLS.

\section{Background}
\label{sec:background}

There has been a broad interest in methods that integrate deep neural nets into iterative algorithms for solving imaging problems. One of the earliest methods is \emph{plug-and-play priors (PnP)}~\cite{Venkatakrishnan.etal2013}, which proposed to extend the traditional proximal optimization~\cite{Parikh.Boyd2014} by replacing the proximal operator with a more general denoiser. It has been shown that the combination of proximal algorithms with advanced denoisers, such as BM3D~\cite{Dabov.etal2008} or DnCNN~\cite{Zhang.etal2017}, leads to the state-of-the-art performance for various imaging problems. Remarkably, the heuristic of using denoisers within iterative algorithms exhibited great empirical success~\cite{Chan.etal2016, Sun.etal2018a, Teodoro.etal2019, Ryu.etal2019, Liu.etal2021b} and inspired a significant follow up work on the so-called \emph{model-based deep learning} method that include RED, \emph{denoising-based approximate message passing (D-AMP)}, \emph{deep unfolding (DU)}, and \emph{deep equilibrium models (DEQ)}~\cite{Yang.etal2016, Metzler.etal2017, Zhang.Ghanem2018, Aggarwal.etal2019, Bertocchi.etal2020, Gilton.etal2021}.

\begin{algorithm}[t]
\caption{RED with BLS}\label{alg:redbls}
\begin{algorithmic}[1]
\State \textbf{input: } $\xbm^0 \in \R^n$; $\gamma, \varepsilon > 0$; $\beta \in (0, 1)$
\For{$k = 1, 2, \dots, t$}
\State $\xbm^k \leftarrow \xbm^{k-1} - \gamma \Gsf(\xbm^{k-1})$
\While{$\|\Gsf(\xbm^k)\|_2 > \|\Gsf(\xbm^{k-1})\|_2$}
\State $\gamma \leftarrow \beta \gamma$
\State $\xbm^k \leftarrow \xbm^{k-1} - \gamma \Gsf(\xbm^{k-1})$
\If{$\gamma < \varepsilon$}
\State \textbf{return:} $\xbmast \leftarrow \xbm^{k-1}$
\EndIf
\EndWhile
\EndFor
\State \textbf{output:} $\xbmast \leftarrow \xbm^t$
\end{algorithmic}
\end{algorithm}%

One of the key motivations of RED is its formulation as an optimization problem~\eqref{Eq:RegularizedInversion} with an explicit regularizer $h$ in~\eqref{Eq:REDRegularizer}. However, the existence of $h$ only holds under a set of assumptions on $\Dsf$, such as its \emph{nonexpansiveness}~\cite{Romano.etal2017, Reehorst.Schniter2019, Cohen.etal2021}. As a reminder, an operator $\Dsf$ is nonexpansive, if it satisfies
\begin{equation}
\|\Dsf(\xbm)-\Dsf(\zbm)\|_2 \leq \|\xbm-\zbm\|_2, \quad \forall \xbm, \zbm \in \R^n.
\end{equation}
Remarkably, nonexpansiveness is also used in the convergence analysis of RED algorithms when interpreting them as fixed-point iterations seeking zeros of the operator $\Gsf$~\cite{Sun.etal2020a, Liu.etal2021b}. While it is possible to train nonexpansive deep neural nets via spectral normalization~\cite{Miyato.etal2018, Ryu.etal2019, Sun.etal2020a}, nonexpansiveness can hurt the performance and reduce the flexibility to use existing expansive denoisers. It is worth noting that nonexpansiveness, as well as more restrictive conditions \emph{firm nonexpansiveness} and \emph{contractiveness}, are used to ensure convergence of other related frameworks, such as PnP and DEQ~\cite{Sun.etal2020a, Gilton.etal2021}. Thus, while our focus is on RED, our approach can be applied to develop more stable variants of other related frameworks.

\section{Proposed Method}
\label{sec:proposed}

\subsection{RED with Backtracking Line Search}

RED algorithms can be interpreted as fixed-point iterations seeking a vector $\xbmast$ in the zero set of the operator $\Gsf$ in~\eqref{Eq:REDSD2}
\begin{equation}
\xbmast \in \zer(\Gsf) \defn \{\xbm \in \R^n : \Gsf(\xbm) = \zerobm\}.
\end{equation}
This implies that the goal of a RED algorithm is to reduce the quantity $\|\Gsf(\xbm^k)\|_2$ at every iteration $k \geq 1$. This observation is behind the \emph{backtracking line search (BLS)} strategy for RED summarized in Algorithm~\ref{alg:redbls}. The main idea behind RED (BLS) is to shrink the step size $\gamma > 0$ by a factor $0< \beta < 1$ whenever the distance to $\zer(\Gsf)$ is not improving. The algorithm stops when the step-size becomes smaller than a pre-defined quantity $\varepsilon > 0$ or when the maximum number of iterations $t \geq 1$ is reached. This strategy was originally proposed in~\cite{Liu.etal2020} as an heuristic to stabilize the convergence of \emph{regularization by artifact removal (RARE)} variant of RED that replaces image denoisers by artifact-removal operators. However, the convergence of the algorithm was never formally compared to the traditional RED with a fixed step-size.

When the data-fidelity term $g$ is a convex Lipschitz continuous function and the denoiser is nonexpansive, one can show that there exists a small enough step size so that the iterates of RED (BLS) satisfy $\|\Gsf(\xbm^k)\|_2 \leq \|\Gsf(\xbm^{k-1})\|_2$ (see the analysis in~\cite{Sun.etal2020a}). However, for nonconvex data-fidelity terms and/or expansive denoisers there is no guarantee for the existence of such a step size, which means that RED (BLS) might terminate prematurely.

\begin{figure}[t]
   \centering
   \includegraphics[width=.9\textwidth]{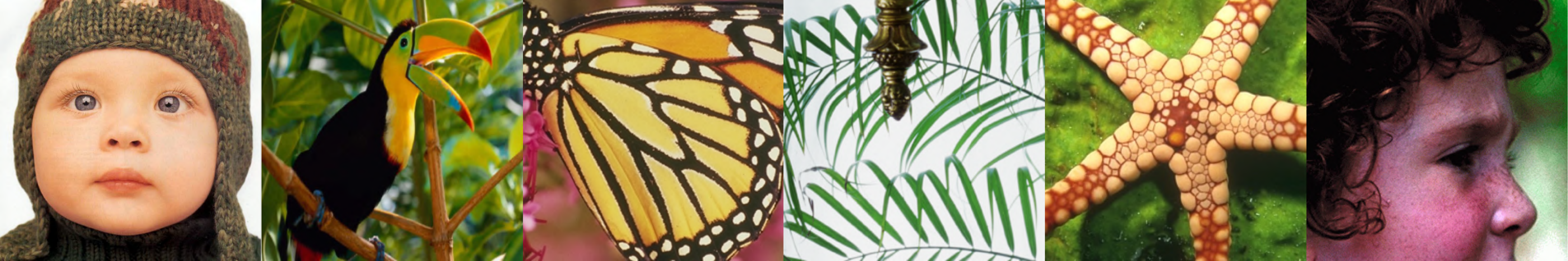}
   \caption{The six test images used in our numerical evaluation. From left to right: \emph{baby, bird, butterfly, leaves, starfish}, and \emph{head}.}
   \label{fig:data}
\end{figure}

\subsection{Monotone RED}

\begin{algorithm}[t]
\caption{Monotone RED (MRED)}\label{alg:mred}
\begin{algorithmic}[1]
\State \textbf{input: } $\xbm^0 \in \R^n$; $\gamma, \alpha_0, \varepsilon > 0$; $\beta \in (0, 1)$; $\theta \in (0, \nicefrac{1}{2})$
\For{$k = 1, 2, \dots, t$}
\State $\xbm^k \leftarrow \xbm^{k-1} - \gamma \Gsf(\xbm^{k-1})$
\State $\alpha \leftarrow \alpha_0$
\While{$\varphi(\xbm^k) > \varphi(\vbm^{k-1})-\alpha\theta\|\nabla \varphi(\xbm^{k-1})\|_2^2$}
\State $\xbm^k \leftarrow \xbm^{k-1} - \alpha \nabla \varphi(\xbm^{k-1})$
\State $\alpha \leftarrow \beta \alpha$
\If{$\alpha < \varepsilon$}
\State \textbf{return:} $\xbmast \leftarrow \xbm^{k-1}$
\EndIf
\EndWhile
\EndFor
\State \textbf{output:} $\xbmast \leftarrow \xbm^t$
\end{algorithmic}
\end{algorithm}%

The proposed \emph{monotone RED (MRED)} method, summarized in Algorithm~\ref{alg:mred}, addresses this issue by reformulating the search for a vector in $\zer(\Gsf)$ as an optimization problem. Consider the loss function
\begin{equation}
\varphi(\xbm) \defn \frac{1}{2}\|\Gsf(\xbm)\|_2^2,
\end{equation}
which measures the squared distance of a vector to $\zer(\Gsf)$. It is clear that whenever $\zer(\Gsf) \neq \varnothing$, the minimizers of $\varphi$ are vectors in $\zer(\Gsf)$. Additionally, whenever $\varphi$ is a smooth function, we can use the traditional optimization theory to perform BLS (see Secion~9.2 in \cite{Boyd.Vandenberghe2004}). To that end, let $\varphi$ be a smooth function that has a Lipschitz continuous gradient with constant $L > 0$ and consider the gradient update
\begin{equation}
\label{Eq:GradUpd}
\xbm^+ \leftarrow \xbm - \alpha \nabla \varphi(\xbm),
\end{equation}
where we will assume that the step satisfies $0 < \alpha \leq 1/L$. Note that one can use the automatic differentiation capability of modern deep learning frameworks to evaluate $\nabla \varphi$. Then, from the Lipschitz continuity of the gradient
\begin{align}
\nonumber\varphi(\xbm^+) &\leq \varphi(\xbm) + \nabla \varphi(\xbm)^\Tsf(\xbm^+-\xbm) + \frac{L}{2}\|\xbm^+-\xbm\|_2^2\\
\nonumber&\leq \varphi(\xbm) - \frac{\alpha}{2} \|\nabla \varphi(\xbm)\|_2^2 \\
\label{Eq:LipUpdBound}&\leq \varphi(\xbm) - \alpha \theta \|\nabla \varphi(\xbm)\|_2^2,
\end{align}
with $0 < \theta < 1/2$, where in the second row we used~\eqref{Eq:GradUpd} and the fact that $0 < \alpha \leq 1/L$. Therefore, there always exists a small enough step-size $\alpha > 0$ such that~\eqref{Eq:LipUpdBound} is satisfied. MRED uses the violation of the condition~\eqref{Eq:LipUpdBound} as an indicator to reduce its step-size parameter. It is worth mentioning that for $\theta = 0$, the condition in Line 5 of MRED is equivalent to the condition of the while loop in Line 4 of RED (BLS). However, unlike RED (BLS), MRED guarantees the existence of a small enough step-size even for nonconvex functions $g$ and expansive denoisers $\Dsf$, so long the function $\varphi$ is smooth.

\begin{figure*}
   \centering
   \includegraphics[width=0.95\textwidth]{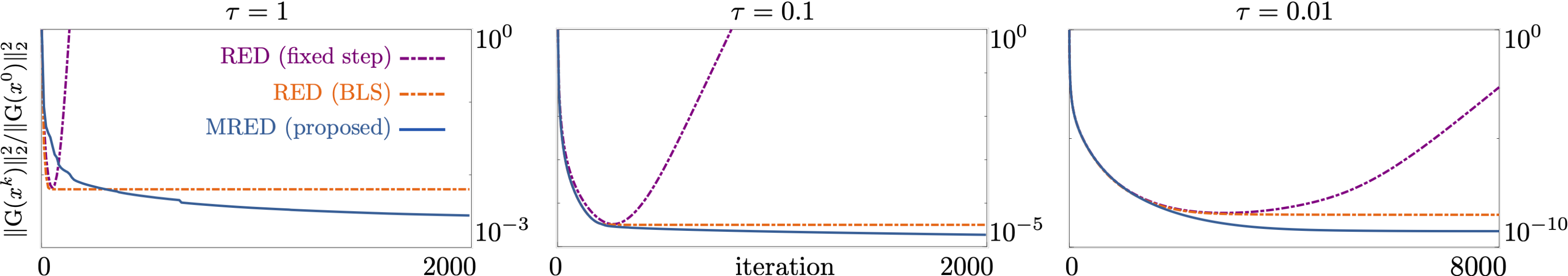}
   \caption{Illustration of the MRED convergence for image deblurring for different regularization parameters ${\tau \in \{1,0.1, 0.01\}}$. The figures plot the average normalized distance to $\zer(\Gsf)$ over the six test images against the iteration number. We compare three algorithms, the traditional RED with a fixed step-size parameter, RED with BLS, and the proposed MRED. Note how MRED achieves the smallest distance to $\zer(\Gsf)$ for all three plots.}
   \label{fig:deblur-conv}
\end{figure*}

\begin{figure*}[htb]
   \centering
   \includegraphics[width=0.95\textwidth]{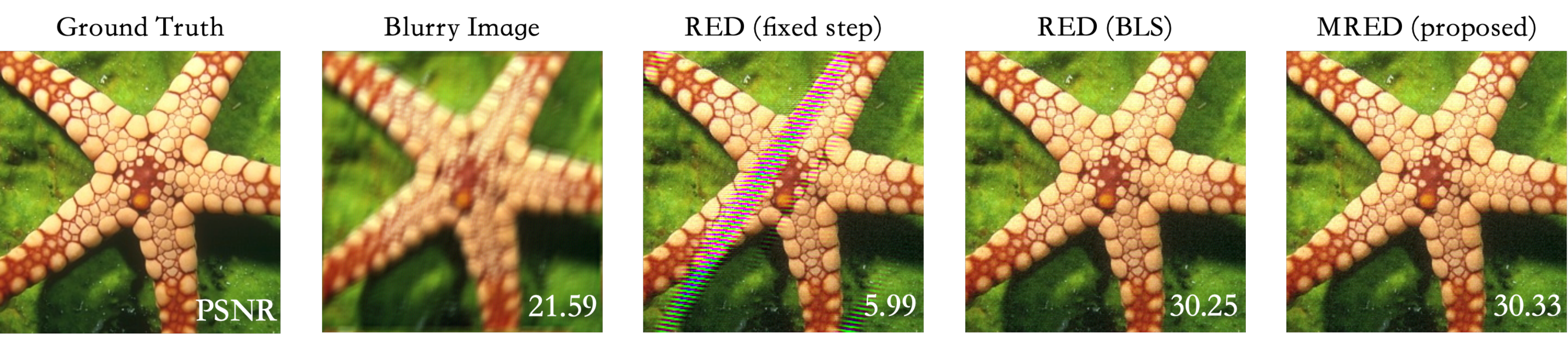}
   \caption{Visual results for image deblurring with a $17 \times 17$ blur kernel from ~\cite{levin2009understanding} and $30 \, \textup{dB}$ AWGN noise on the \emph{Starfish} image. We ran three algorithms the traditional RED with the fixed step-size, RED (BLS), and MRED for the regularization parameter $\tau=0.01$. Note the poor visual quality of the RED image with the fixed step-size due to divergence of the algorithm.}
   \label{fig:deblur-visual}
\end{figure*}

MRED is a hybrid algorithm in the sense that it uses the traditional RED update (Line 3 of MRED) whenever it can reduce the distance to $\zer(\Gsf)$ and switches to the gradient update (Line 6 of MRED) if this is not possible. The benefit of the proposed hybrid approach is that it maintains the fast fixed-point convergence of the traditional RED algorithm whenever possible. Additionally, the gradient $\nabla \varphi$ needs to be evaluated only when the traditional RED update cannot reduce the loss, thus limiting the computational overhead relative to the traditional RED algorithm. In short, MRED maintains all the benefits of the traditional RED algorithm, while also providing additional flexibility and stability. In Section~\ref{sec:numerical}, we validate the effectiveness of MRED by comparing it both with the traditional RED with a fixed step-size and RED (BLS) on two different types of inverse problems. 

\section{Numerical Results}
\label{sec:numerical}

\begin{figure*}[htb]
  \centering
  \includegraphics[width=0.95\textwidth]{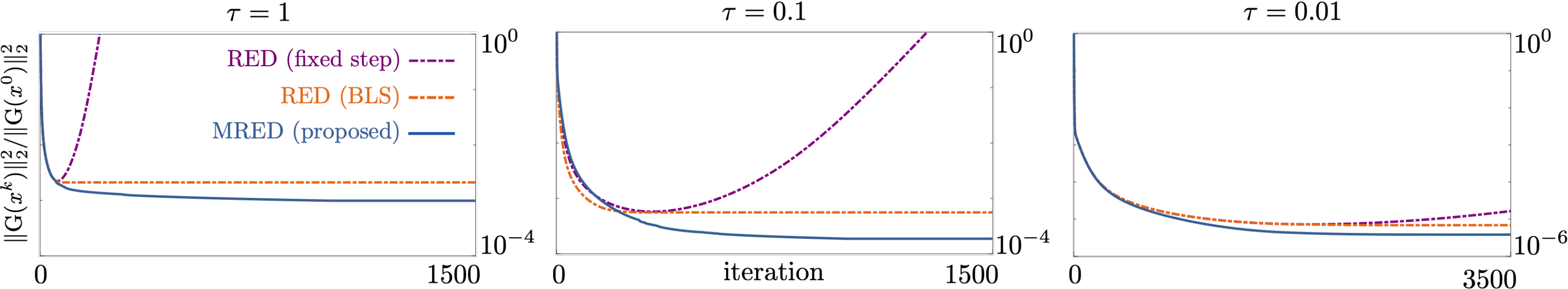}
  \caption{Illustration of the MRED convergence for compressive sensing from random measurements for different regularization parameters ${\tau \in \{1,0.1, 0.01\}}$. The figures plot the average normalized distance to $\zer(\Gsf)$ over the six test images against the iteration number. We compare three algorithms, the traditional RED with a fixed step-size parameter, RED with BLS, and the proposed MRED. Note how MRED achieves the best convergence behaviour for all three plots.}
  \label{fig:cs-conv}
\end{figure*}

In this section, we numerically illustrate the ability of MRED to stabilize the convergence for expansive deep denoising priors. We consider two scenarios: (a) image deblurring and (b) compressive sensing using random projections. Our deep neural net prior is the simplified Unet architecture~\cite{Ronneberger.etal2015}, obtained by removing all the batch-normalization layers. We trained the denoiser as a residual network ${\Rsf \defn \Isf - \Dsf}$ that predicts the noise residual from a noisy input, without using any spectral normalization for controlling the nonexpansiveness of the network. Hence, our network is expected to be expansive. The training data was generated by adding AWGN to images from the BSD500~\cite{Martin.etal2001} and DIV2K~\cite{agustsson2017ntire} datasets, by cropping the images into small patches of $96 \times 96$ pixels with stride 10. Note that the gradient in~\eqref{Eq:GradUpd} is given by the expression $\nabla \varphi(\xbm) = \Abm^\Tsf\Abm\Gsf(\xbm) + \tau\nabla\Rsf(\xbm)^\Tsf\Gsf(\xbm)$, where the last term was computed by using the autodiff functionality of PyTorch. The normalized distance $\|\Gsf(\xbm^k)\|_2^2 /\|\Gsf(\xbm^{0})\|_2^2$ was used to quantify the fixed-point convergence of three RED algortihms: (a) the traditional RED with a fixed step size; (b) RED with BLS in Algorithm~\ref{alg:redbls}; and (c) MRED in Algorithm~\ref{alg:mred}. Following the theoretical analysis in~\cite{Sun.etal2020a}, we fix the step-size of the traditional RED to $\gamma = 1/(L+2\tau)$, which is small enough to theoretically ensure convergence under nonexpansive denoisers. Note that we fixed The quantity $\|\Gsf(\xbm^k)\|_2^2 /\|\Gsf(\xbm^{0})\|_2^2$ is expected to approach zero as the algorithms converge to $\zer(\Gsf)$. All the quantitative results were obtained by averaging over six test images shown in Fig.~\ref{fig:data}. 

\subsection{Image Deblurring}

The measurement model in image deblurring with uniform blur can be expressed as in~\eqref{Eq:ForwardProblem}, where the matrix $\Abm$ denotes a two-dimensional convolution between the clean image $\xbm$ and the blur kernel. We consider the blur kernel of size $17 \times 17$ and the AWGN vector $\ebm$ corresponding to the input SNR of $30 \, \textup{dB}$. Fig.~\ref{fig:deblur-conv} illustrates the convergence behaviour of all three RED algorithms. First, note the divergence of the traditional RED with the fixed step-size, which is not surprising since our CNN prior has been trained without any Lipschitz constraints. Note how RED (BLS) stabilizes the convergence, but stops making any progress prematurely due to the fact that it cannot find a small enough step-size to decrease the distance to $\zer(\Gsf)$. Finally, as expected, MRED achieves the best convergence behaviour for all three values of $\tau$. Fig.~\ref{fig:deblur-visual} shows some visual examples of the performance of all three algorithms, where RED yields a suboptimal image due to its divergence and MRED achieves the highest quality image.



\subsection{Compressive Sensing}

We consider the traditional compressive sensing measurement setup $\ybm = \Abm\xbm$, where $\Abm \in \R^{m \times n}$ is orthogonalized version of a matrix with elements that are i.i.d. zero-mean Gaussian random variables of variance of $1/m$. We set the compression ratio $m/n$ to be $0.1$ and consider a similar setting to image deblurring by running all three algorithms for three values of $\tau$: 1, 0.1, and 0.01. As before, our baseline methods are the traditional RED with a fixed step-size and RED (BLS). Fig.~\ref{fig:cs-conv} illustrates the divergence of the traditional RED and convergence of MRED for all the regularization parameters $\tau$.

\section{Conclusion}
\label{sec:conclusion}

There is a growing interest in iterative algorithms, such as PnP, RED, or DEQ, for solving inverse problems by using deep neural nets as priors. While such algorithms have often lead to significant improvements in the imaging quality, they have also been shown to suffer from convergence issues when the deep neural net is expansive. In this paper, we proposed a new MRED algorithm that uses a novel line search strategy to ensure monotonic convergence of the steepest descent variant of RED (RED-SD) even for nonconvex data-fidelity terms and expansive denoisers. Our numerical results highlight the stability of MRED relative to the traditional RED-SD algorithm, which diverges for our expansive denoiser. While our focus was on RED, the future work will look into the potential of our strategy to be applicable to other related frameworks such as PnP and DEQ.

\section*{Acknowledgement}
This work was supported by the NSF award CCF-2043134.

\bibliographystyle{IEEEbib}

\end{document}